\documentclass[a4paper,USenglish]{lipics-v2021}

\usepackage{amsmath,amssymb}
\usepackage{algorithm}
\usepackage{algpseudocode}
\usepackage{tikz}
\usetikzlibrary{positioning}
\usepackage[section]{placeins} % keep each algorithm inside the section that introduces it

\newcommand{\copfill}{\operatorname{copfill}}
\newcommand{\fillin}{\operatorname{fill}}
\newcommand{\nd}{\operatorname{nd}}
\newcommand{\vc}{\operatorname{vc}}
\newcommand{\tw}{\operatorname{tw}}
\newcommand{\supp}{\operatorname{supp}}
\newcommand{\problem}[1]{\textnormal{\textsc{#1}}}

\title{Cop-Win Edge Completion: Hardness, Structure, and Parameterized Algorithm}
\titlerunning{Cop-Win Edge Completion}
\author{Rajat Adak}{Department of Computer Science and Automation, Indian Institute of Science, Bangalore, India}{rajatadak@iisc.ac.in}{}{}
\authorrunning{R. Adak}
\Copyright{Rajat Adak}
\hideLIPIcs
\nolinenumbers
\hypersetup{pdfauthor={Rajat Adak},pdftitle={Cop-Win Edge Completion: Hardness, Structure, and Parameterized Algorithm}}

\begin{document}
\maketitle
\begin{abstract}
We study \textsc{Cop-Win Edge Completion}, the problem of determining the minimum number of edges that must be added to a graph so that one cop can capture a visible robber.

We first give an exact reduction from \textsc{Dominating Set}. For every nonempty graph $H$, we construct a graph $G_H$ on $2|V(H)|+2$ vertices such that \[\copfill(G_H)=\gamma(H).\]

This resolves a complexity question posed by Mertzios et al. and implies NP-completeness even for connected graphs of diameter two with a dominating edge and domination number two. The problem is also $\mathrm{W}[2]$-hard parameterized by the completion budget $k$, and even by $k+\gamma(G)$.

This gives an interesting contrast with the classical \textsc{Minimum Fill-in}
problem. Every connected chordal graph is cop-win, and the inclusion is strict.
Thus, on connected inputs, cop-win completion allows a strictly larger target
class than chordal completion. Interestingly, while \textsc{Minimum Fill-in} is
fixed-parameter tractable with respect to the number of added edges,
\textsc{Cop-Win Edge Completion} is $\mathrm{W}[2]$-hard for the same
parameter.

We then give two lower bounds for unrestricted cop-win completion. The first shows that $\copfill(G)\ge2n-m-b-c-1$, where $n,m,b,c$ denote the numbers of vertices, edges, nontrivial blocks, and connected components. It is tight for partial $2$-trees, giving a linear-time optimum value algorithm and a quadratic-time construction. For outerplanar graphs, an optimum completion can be constructed in linear time while preserving outerplanarity. The second bound uses the cycle space modulo the subspace generated by triangles. It gives the exact formula,
 $\copfill(G)=c-1+\sum_{i=1}^r(\ell_i-3)$,
for block-cactus graphs with long cycle blocks of lengths $\ell_1,\ldots,\ell_r$.

We also show how the optimum behaves under disjoint unions and joins. These results give a polynomial-time algorithm for cographs and, more generally, $P_4$-sparse graphs, and exact formulas for complete multipartite graphs. In contrast to partial $2$-trees and block-cactus graphs, an optimum cop-win completion can be much smaller than a minimum chordal completion, already on complete bipartite graphs.

Finally, we give fixed-parameter algorithms for $k+\nd(G)$, for the twin-cover number $t=\operatorname{tc}(G)$ alone in time $2^{O(t^3)}n^{O(1)}$, and for the combined parameter $k+\tw(G)$.
\end{abstract}
\clearpage

\section{Introduction}\label{section_intro}
The game of Cops and Robbers is a classical game played on graphs. A cop and a robber occupy vertices of the graph and move along its edges, with the cop seeking to capture the robber by moving to the vertex occupied by the robber. A graph is \emph{cop-win} if one cop can capture a visible robber in the standard alternating-move pursuit game. In \cite{nowakowski1983vertex}, the finite cop-win graphs were characterized to be exactly the \emph{dismantlable} graphs in which a vertex whose closed neighborhood is contained in that of another vertex can be deleted repeatedly until one vertex remains. This characterization makes recognition polynomial-time. We ask instead how many edges must be added to obtain such a graph. For a nonempty graph $G$, define
\[
 \copfill(G)=\min\bigl\{|F|:F\subseteq\tbinom{V(G)}2\setminus E(G),\ G+F\text{ is cop-win}\bigr\}.
\]
The decision problem \problem{Cop-Win Edge Completion} asks whether $\copfill(G)\le k$.

The problem arises explicitly in the recent paper of Mertzios, Michail, Skretas, Spirakis, and Theofilatos~\cite{mertzios2025complexity}. They relate cop-win completion to minimizing excess edges when the number of growth steps is unrestricted, and ask whether this optimization problem is polynomial-time solvable or NP-hard. We resolve this question by an optimum-preserving reduction with two auxiliary vertices.

\paragraph*{Comparison with chordal completion}
The classical \problem{Minimum Fill-in} problem asks for the fewest edges needed to make a graph chordal. Write its optimum as $\fillin(G)$. Its decision version is NP-complete~\cite{yannakakis1981computing}, but FPT for the number of added edges~\cite{kaplan1999tractability}. Fomin and Villanger~\cite{fomin2013subexponential} gave an algorithm running in $O(2^{O(\sqrt{k}\log k)}+k^2nm)$ time. Every connected chordal graph is cop-win, so for connected inputs
\[
 \copfill(G)\le\fillin(G).
\]
The target cop-win class is strictly larger, since a wheel is cop-win but need not be chordal. However, our result shows that cop-win completion is $\mathrm{W}[2]$-hard for $k$. We also study when chordal completion is still optimal. The lower bounds must allow edges between different blocks or components, rather than assuming that the completion respects the input decomposition.

\paragraph*{Results and proof ideas}
Our first result is an exact reduction from \problem{Dominating Set}. For every nonempty graph $H$, we construct a graph $G_H$ such that
$\copfill(G_H)=\gamma(H)$. A dominating set gives a completion that the cop can win in two moves. In the other direction, we use a retraction onto a four-cycle and a counting argument to obtain a dominating set from any completion. The constructed graphs have diameter two and domination number two. This proves NP-completeness and $\mathrm{W}[2]$-hardness for $k$ and for $k+\gamma(G)$.

We then study when an optimum chordal completion is also an optimum cop-win completion. The first lower bound is $\copfill(G)\ge2n-m-b-c-1$, where $b$ counts blocks with an edge and $c$ counts components. This bound is tight for partial $2$-trees. It gives a linear-time optimum value algorithm and a quadratic-time construction, while outerplanar graphs also admit a linear-time optimum construction. Our second bound uses cycles that are not sums of triangles. It gives the exact value $c-1+\sum_i(\ell_i-3)$ on block-cactus graphs, together with a linear-time construction.

We also show how the optimum behaves under disjoint unions and joins. They give linear-time algorithms for cographs and $P_4$-sparse graphs. Complete bipartite graphs then show that an optimum cop-win completion can be much smaller than an optimum chordal completion. The two optima differ already for $K_{3,3}$, which has treewidth three.

Finally, we use symmetry to obtain FPT algorithms for $k+\nd(G)$ and for the twin-cover number $t=\operatorname{tc}(G)$, the latter in $2^{O(t^3)}n^{O(1)}$ time. An orientation of the completed graph also gives FPT for $k+\tw(G)$.

\paragraph*{Related context and Organization}
Cai's framework~\cite{cai1996fixed} gives FPT algorithms for graph modification
to hereditary classes described by finitely many forbidden induced subgraphs. It does not apply directly to cop-win graphs, because the cop-win property is not hereditary. Cai also showed separately that chordal completion is FPT when parameterized by the number of added edges. In contrast, we show that \problem{Cop-Win Edge Completion} is $\mathrm{W}[2]$-hard for the same parameter.

Our arguments use dismantling orders, retractions, cycle-space arguments, and symmetry.
Section~\ref{section_prelim} gives the preliminaries and  notations.
Section~\ref{section_hardness} presents the exact reduction and its complexity consequences.
Sections~\ref{section_lower} and~\ref{section_copfill-block-cactus} develop the two structural lower bounds and their algorithmic applications.
Section~\ref{section_composition} establishes the disjoint-union and join formulas and applies them to cographs and complete multipartite graphs.
Section~\ref{section_direct-fpt} presents the parameterized algorithms.
Recognition procedures, explicit completion algorithms and their implementation details, and the formal logical encoding used by the treewidth algorithm are given together with the corresponding results.

\section{Preliminaries}\label{section_prelim}
All graphs are finite, simple, undirected, and nonempty unless stated otherwise. Put $n=|V(G)|$, $m=|E(G)|$, and let $c=c(G)$ be the number of connected components. Write $N_G(v)$ and $N_G[v]$ for open and closed neighborhoods, and $N_G[D]=\bigcup_{v\in D}N_G[v]$. A dominating set $D$ satisfies $N_G[D]=V(G)$. The minimum size of a dominating set is $\gamma(G)$, the domination number. A vertex cover is a subset of $V(G)$ meeting every edge, and its minimum size is $\vc(G)$. An edge $ab$ is \emph{dominating} if $N_G[a]\cup N_G[b]=V(G)$.

The cop chooses a starting vertex, then the robber chooses one, and the cop moves first. On each turn a player may stay or traverse an edge. Capture occurs when the cop arrives at the vertex currently occupied by the robber, and the cop wins. A vertex $v$ is \emph{dominated} by a distinct vertex $w$ if $N[v]\subseteq N[w]$. A \emph{dismantling order} deletes dominated vertices until one remains. From \cite{nowakowski1983vertex}, existence of a dismantling order in a graph is equivalent to it being cop-win. In particular, a cop-win graph is connected. Clearly a connected chordal graph is cop-win, a simplicial vertex exists~\cite{dirac1961rigid}, any of its neighbors dominates it, and its deletion preserves connectivity and chordality. Cop-win recognition can be done in $O(n^3)$ time, as explained at the end of this section.

A \emph{retraction} from $J$ onto a subgraph $R$ fixes $R$ pointwise and maps adjacent vertices to equal or adjacent vertices. It is easy to check that retracts of cop-win graphs are cop-win. Simulate a winning strategy in $J$ against a robber in $R$ and project the cop's positions. We use this only for explicit maps.

Let $\mathcal B(G)$ contain the maximal $2$-connected subgraphs of order at least three and one $K_2$ block for each bridge. Isolated vertices are not assumed to be blocks. Write $b=b(G)=|\mathcal B(G)|$ and $n_B=|V(B)|$, $m_B=|E(B)|$. We can use depth-first search to find components and blocks in $O(n+m)$ time~\cite{tarjan1972depth}.

For disconnected inputs, the minimum size of a \emph{connected chordal} completion is $\fillin(G)+c-1$. Restriction to each original component requires its own chordal fill, and connectivity requires at least $c-1$ intercomponent edges. Conversely, join optimum chordal completions of the components by bridges. Thus $\copfill(G)\le\fillin(G)+c-1$ for every input.

\paragraph*{Cubic-time cop-win test}
A state describes the current position of the game. We use the
$2n^2$ states $(c,r,\sigma)$, where $c$ is the cop's vertex, $r$ is
the robber's vertex, and $\sigma$ tells whose turn it is. On a move,
the active player may stay at the same vertex or move to a neighbor,
and then the turn changes.

 Call a state winning if the cop can force a capture from that state. Every state with $c=r$ is winning. We then work backwards. A cop-turn state is winning if the cop has some move to a winning state. A
robber-turn state is winning if every move of the robber leads to a
winning state. We apply these two rules repeatedly until no new winning states are found.
The states found in this way are exactly those from which the cop can
force a capture.

These rules are correct because from a winning cop-turn state, the cop
can choose a move that leads to another winning state, while from a
winning robber-turn state, every robber move leads to a winning state.
If a state is not winning, then the robber can always choose moves that
avoid the winning states and therefore avoid capture.

The winning states can be found efficiently using a queue. For each
state, store the states that can move to it. For each robber-turn state,
also keep track of how many of its possible moves are not yet known to
be winning. Each state and each transition is processed only once.
Since there are $O(n^3)$ transitions, the algorithm takes $O(n^3)$ time
and space.

Finally, $G$ is cop-win exactly when there is a starting vertex $c$ for
the cop such that $(c,r,\text{cop})$ is winning for every possible
starting vertex $r$ of the robber. 

\section{Exact hardness reduction}\label{section_hardness}
We reduce from \problem{Dominating Set}. For each vertex $u\in V(H)$,
the construction contains a vertex $s_u$, and choosing $u$ for the
dominating set corresponds to adding the edge $bs_u$. This edge allows
$s_u$ to dominate every vertex $x_v$ with $v\in N_H[u]$.
The main difficulty is the reverse direction, since a completion may
also use other missing edges.

\subsection{The construction}
Let $H$ be a nonempty graph. For each $u\in V(H)$, introduce two vertices $s_u$ and $x_u$, and put
\[
 S=\{s_u:u\in V(H)\},\qquad X=\{x_u:u\in V(H)\}.
\]
Add two more vertices $a,b$ and set $V(G_H)=S\cup X\cup\{a,b\}$. 

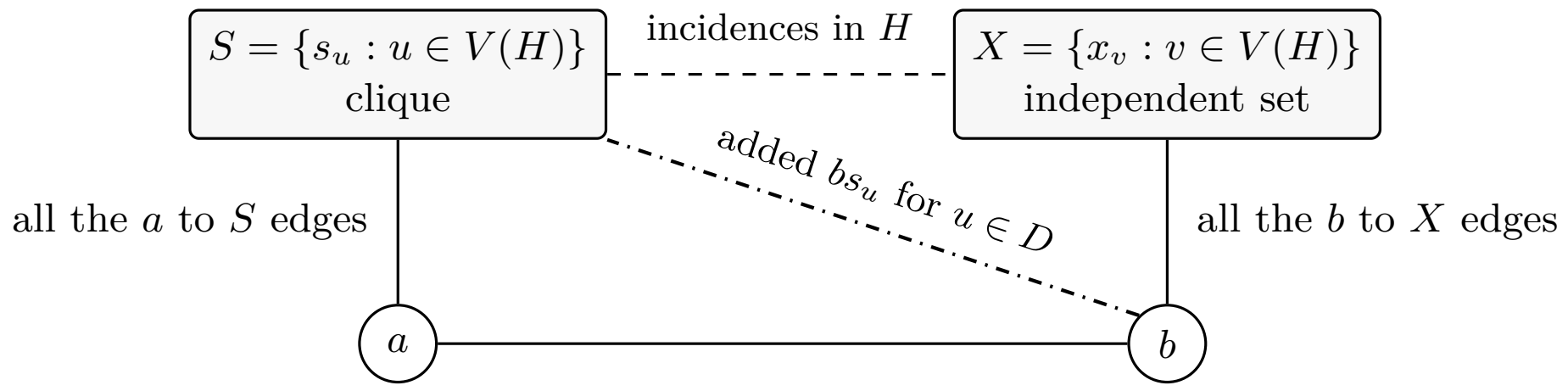
\begin{figure}[htbp]
 \centering
    \begin{tikzpicture}[
  vertex/.style={circle,draw,fill=white,minimum size=6mm,inner sep=1pt},
  class/.style={draw,rounded corners=2pt,fill=black!3,align=center,
                minimum width=29mm,minimum height=10mm,inner sep=4pt},
  every node/.style={font=\small},
  line width=0.6pt
]
  \node[class] (S) at (0,2.1) {$S=\{s_u:u\in V(H)\}$\\clique};
  \node[class] (X) at (6,2.1) {$X=\{x_v:v\in V(H)\}$\\independent set};
  \node[vertex] (a) at (0,0) {$a$};
  \node[vertex] (b) at (6,0) {$b$};
  \draw (a) -- node[left=3pt] {all the $a$ to $S$ edges} (S);
  \draw (b) -- node[right=3pt] {all the $b$ to $X$ edges} (X);
  \draw (a) -- (b);
  \draw[dashed] (S.east) -- node[above=4pt,font=\footnotesize] {incidences in $H$} (X.west);
  \draw[dash dot,line width=0.8pt] (b.north west) --
    node[sloped,above=4pt,font=\footnotesize,fill=white,inner sep=1pt]
    {added $bs_u$ for $u\in D$} (S.south east);
\end{tikzpicture}
 \caption{The graph $G_H$. The dashed line denotes the closed-neighborhood incidences $s_ux_v$ with $v\in N_H[u]$. The dash-dotted diagonal represents the added fill-edges $bs_u$, for all $u\in D$}
 \label{redution_figure}
\end{figure}

Make $S$ a clique and $X$ an independent set. Join $a$ to every vertex of $S$, join $b$ to every vertex of $X$, and add $ab$. The remaining edges are exactly
\[ s_ux_v\in E(G_H)\quad\iff\quad v\in N_H[u].\]
Note that in $G_H$, $a$ is not adjacent to any vertex in $X$, and $b$ is not adjacent to any vertex in $S$.  Clearly, $G_H$ has $2|V(H)|+2$ vertices and includes every edge $s_ux_u$. For $D\subseteq V(H)$, let $F_D=\{bs_u:u\in D\}$. Figure~\ref{redution_figure} shows the edges in $G_H$, and those in $F_D$.

\subsection{Correctness}
\begin{lemma}\label{forward_lemma}
If $D$ dominates $H$, then $G_H+F_D$ is cop-win, with capture within two cop moves. Consequently, $\copfill(G_H)\le\gamma(H)$.
\end{lemma}
\begin{proof}
The cop starts at the vertex $a$. If the robber starts at $a$ then he is already captured, and if he starts at $b$ or in $S$ then he is captured on the first move. Otherwise the robber starts at $x_v$. Choose $u\in D$ with $v\in N_H[u]$ (since $D$ is a dominating set, such a $u$ exists), and move the cop to $s_u$ on the first move. The robber can stay at $x_v$, move to $b$, or move to some $s_w$. The cop at $s_u$ sees $x_v$ by its choice, sees $b$ through $bs_u\in F_D$, and sees all other vertices of $S$ because $S$ is a clique. Thus, every surviving robber position is captured on the next cop move.
\end{proof}

\begin{lemma}\label{C4_lemma}
Let $J$ contain a four-cycle $C$. Suppose there is a map $\rho:V(J)\to V(C)$ that fixes $C$ pointwise and maps every edge of $J$ either to an edge of $C$ or to a single vertex of $C$. Then $J$ is not cop-win.
\end{lemma}
\begin{proof}
The map $\rho$ is a retraction of $J$ onto $C$ in the sense defined in
Section~\ref{section_prelim}. If $J$ were cop-win, then its retract $C$ would
also be cop-win.

However, a four-cycle is not cop-win. Indeed, no vertex of $C_4$ is
dominated by another vertex, so $C_4$ does not have a dismantling order. Hence $C$ is not cop-win, a contradiction.
\end{proof}
For the reverse direction, fix any $F$ such that $J=G_H+F$ is cop-win. Let
\[
 D_0=\{u:bs_u\in F\},\qquad S_0=\{s_u:u\in D_0\},\qquad Q=J[X].
\]
Note that $D_0 \subseteq V(H)$ and can be empty and, $S_0 \subset V(G_H)$. Every edge of $Q$ is a fill edge, since $X$ is an independent set in $G_H$. Say that $x_v$ is \emph{covered} if $v\in N_H[D_0]$. Note that coverage refers only to the original incidences, not to newly added edges between $S$ and $X$. If a component $R$ of $Q$ contains no
covered vertex, then we will see that, at least one additional fill edge outside $Q$ must
be incident with a vertex of $R$.

\begin{lemma}\label{impedge_lemma}
Let $R$ be a component of $Q$ containing no covered vertex. Then $F$ contains an edge $ax_v$ with $x_v\in R$, or an edge $s_ux_v$ with $u\in D_0$ and $x_v\in R$. Such an edge in $F$ is called an \emph{important edge} for $R$.
\end{lemma}
\begin{proof}
Suppose there is no important edge, for $R$. Let $x_v\in R$, therefore $v$ is not covered. Since $v\notin N_H[D_0]$, we have $v\notin D_0$. The vertices $a,b,x_v,s_v$ induce a four-cycle in this order. Its diagonals $ax_v$ and $bs_v$ are absent since we assumed there is no important edge and by $v\notin D_0$, respectively. Define $\rho:V(J)\to V(C)$ by
\[
\rho(z)=
\begin{cases}
a, & z\in \{a\}\cup S_0\cup(X\setminus R),\\
b, & z=b,\\
x_v, & z\in R,\\
s_v, & z\in S\setminus S_0.
\end{cases}
\]
The map $\rho$ fixes the four-cycle, $C=a-b-x_v-s_v-a.$

We now check that every edge of $J$ is mapped either to an edge of $C$
or to a single vertex of $C$.
There is no edge from $b$ to a vertex of $S\setminus S_0$, by the
definition of $D_0$. Also, there is no edge from a vertex of $R$ to
$a$, since such an edge would be an important edge. There is no edge from $R$
to $S_0$: no such original edge exists because $R$ contains no covered
vertex, and no such added edge exists because we assumed that $R$ has
no important edge. Finally, there is no edge from $R$ to $X\setminus R$, since
$R$ is a connected component of $Q=J[X]$.

Thus every edge of $J$ is mapped by $\rho$ either to an edge of $C$ or
to one vertex of $C$. Hence $\rho$ is a retraction of $J$ onto $C$.
By Lemma~\ref{C4_lemma}, $J$ is not cop-win, a
contradiction.
\end{proof}

\begin{lemma}\label{backward_lemma}
From every cop-win completion $F$ of $G_H$, a dominating set of $H$ of size at most $|F|$ can be constructed in polynomial time.
\end{lemma}
\begin{proof}
Set $D=D_0\cup\bigl(V(H)\setminus N_H[D_0]\bigr).$
Then $D$ dominates $H$, since every vertex not already dominated by
$D_0$ is included in $D$. We will show that $|D|\le |F|$.

Recall that $x_v$ is covered if $v\in N_H[D_0]$. For a component $R$ of $Q$, let $t_R$ be the number of covered
vertices in $R$. The remaining $|R|-t_R$ vertices $x_v\in R$
are uncovered, and hence their corresponding vertices $v$ belong to
$D\setminus D_0$.

First suppose that $t_R\ge 1$. Since $R$ is connected, it contains a
spanning tree with $|R|-1$ edges. Every edge of $Q$ is a fill edge, so
these are $|R|-1$ distinct edges of $F$. Since, $|R|-t_R\le |R|-1,$
there are enough such fill edges to account for all vertices of
$D\setminus D_0$ corresponding to $R$.

Now suppose that $t_R=0$. Then all $|R|$ vertices corresponding to
$R$ belong to $D\setminus D_0$. Again, a spanning tree of $R$
contains $|R|-1$ edges of $F$. In addition,
Lemma~\ref{impedge_lemma} gives an important edge for $R$, which is another edge
of $F$, distinct from the $|R|-1$ spanning tree edges. Hence in this case $R$ gets at least $|R|$ distinct edges of
$F$.

The edges counted for different components of $Q$ are distinct.
Moreover, an important edge is not an edge of $Q$, and important edges belonging to
different components are distinct because their endpoints in $X$ lie in
different components.

Finally, for every $u\in D_0$, the edge $bs_u$ belongs to $F$.
These $|D_0|$ edges are distinct from all the edges counted above.
Since the components of $Q$ partition $X$, every vertex of
$D\setminus D_0$ corresponds to exactly one uncovered vertex $x_v$,
and hence
\[
|D\setminus D_0|=\sum_R(|R|-t_R).
\]
Together with the distinct fill edges identified above, this gives
\[
|F|
\ge
|D_0|+\sum_R(|R|-t_R)
=
|D|.
\]
Thus $D$ is a dominating set of $H$ of size at most $|F|$.
\end{proof}

\begin{theorem}\label{hardness_thm}
For every nonempty graph $H$, $\copfill(G_H)=\gamma(H)$. Thus $(H,k)\mapsto(G_H,k)$ is an optimum-preserving, budget-preserving polynomial-time reduction from \problem{Dominating Set} to \problem{Cop-Win Edge Completion}.
\end{theorem}
\begin{proof}
From Lemma~\ref{forward_lemma} we have $\copfill(G_H)\le \gamma(H).$ Conversely, in Lemma~\ref{backward_lemma} if we take $F$ to be a minimum cop-win completion of $G_H$, we get $\gamma(H)\le |F|=\copfill(G_H).$
Therefore $\copfill(G_H)=\gamma(H)$.

The construction of $G_H$ is polynomial-time and leaves the budget
unchanged. Thus $(H,k)$ is a yes-instance of \problem{Dominating Set}
if and only if $(G_H,k)$ is a yes-instance of
\problem{Cop-Win Edge Completion}.
\end{proof}

As a consequence of Theorem~\ref{hardness_thm} we get additional hardness results.

\begin{corollary}\label{hardness_consequences}
\problem{Cop-Win Edge Completion} is NP-complete even for connected diameter-two graphs with a dominating edge and domination number two. It is $\mathrm{W}[2]$-hard for $k$ and for $k+\gamma(G)$, and is para-NP-hard for $\gamma(G)$ alone, under these restrictions. Moreover, the reduction of Theorem~\ref{hardness_thm} is approximation-preserving.
\end{corollary}
\begin{proof}
The edge $ab$ dominates $G_H$, so $G_H$ is connected and $\gamma(G_H)\le2$. Neither $a$ nor $b$ is universal, and no vertex of $S\cup X$ is universal, so $\gamma(G_H)=2$. Every pair of nonadjacent vertices is at distance two. Two vertices of $X$ have the common neighbor $b$, and the vertices $a$ and $b$ reach $X$ and $S$, respectively, in two steps. If $s_u$ and $x_v$ are nonadjacent, then $s_u,s_v,x_v$ is a path of length two. Thus $G_H$ has diameter two.

    Theorem~\ref{hardness_thm} gives NP-hardness. The problem is also in NP, since a proposed completion can be checked in polynomial time. The reduction preserves $k$ and produces graphs with $\gamma(G_H)=2$. Thus, the $\mathrm{W}[2]$-hardness of \problem{Dominating Set}~\cite{downey1995fixed} carries over to both $k$ and $k+\gamma(G)$. It also gives para-NP-hardness for $\gamma(G)$ alone, since this parameter is always two in the constructed graphs.

     Finally, by Lemma~\ref{backward_lemma}, any completion $F$ of $G_H$ can be converted into a dominating set $D$ of $H$ with $|D|\le |F|$. If $F$ is an $\alpha$-approximate completion, then Theorem~\ref{hardness_thm} gives $|D|\le |F|\le\alpha\copfill(G_H)=\alpha\gamma(H).$ Hence, $D$ is also an $\alpha$-approximate dominating set.
\end{proof}

\section{When chordal completion is optimal}\label{section_lower}
We now focus on when chordal completion of a graph is an optimal cop-win completion as well. The first lower bound counts the edges in cop-win blocks and tracks how adding edges changes the block structure.

\subsection{A density lower bound via block structure}
Recall that $\mathcal{B}(G)$ is the set of blocks of $G$, that is, the maximal $2$-connected subgraphs of order at least three, together with one $K_2$ block for each bridge. For $B\in\mathcal{B}(G)$, let $n_B=|V(B)|$ and $m_B=|E(B)|$. Let $c$ be the number of components of $G$.
\begin{equation}\label{block_identities}
 \sum_{B\in\mathcal B(G)}(n_B-1)=n-c,\qquad
 \sum_{B\in\mathcal B(G)}m_B=m.
\end{equation}
For the first identity, within each connected component, start with any block and add the remaining blocks one at a time. Each new block shares exactly one vertex with the previously added blocks, and hence contributes $n_B-1$ new vertices. Summing over all components gives $n-c$. The second identity follows because every edge belongs to exactly one block. An isolated vertex contributes zero to both sums.

Note that every block of a connected graph is a retract. This is because each component outside the block attaches at exactly one block vertex and the whole component can be mapped there. If there are two distinct attachment vertices then we will get a contradiction to the maximality of the block. Therefore, every block of a cop-win graph is cop-win.

\begin{lemma}\label{block_monotonicity}
If $J$ is a connected spanning supergraph of $G$, then $b(J)\le b(G)+c(G)-1$.
\end{lemma}
\begin{proof}
    Insert the new edges of $J$ one at a time. If an added edge has both endpoints
in the same connected component, then it either stays within one block
or creates a cycle through several blocks, causing those blocks to merge.
Thus the number of blocks does not increase.

If an added edge joins two different connected components, it forms a new
bridge block and increases the number of blocks exactly by one. Since exactly
$c(G)-1$ added edges are needed to make the graph connected, we get,
$b(J)\le b(G)+c(G)-1.$
\end{proof}

\begin{lemma}\label{copfill_density}
A $2$-connected cop-win graph $J$ on $s\ge3$ vertices has at least $2s-3$ edges. Consequently, every connected cop-win graph satisfies
\begin{equation}\label{copfill_density_con}
 |E(J)|\ge 2|V(J)|-b(J)-2.
\end{equation}
\end{lemma}
\begin{proof}
We prove the first statement by induction on $s$, with $K_2$ as the
base case. Let $J$ be a $2$-connected cop-win graph on $s\ge 3$
vertices. Since $J$ is cop-win, there is a vertex $v$ dominated by
some vertex $w$. Let $J'=J-v.$
Then $J'$ is connected and cop-win.

We first claim that no vertex other than $w$ can be a cutvertex of
$J'$. Suppose that some $x\ne w$ is a cutvertex. Since $v$ is
dominated by $w$, every neighbor of $v$ that remains in $J'-x$ is
either $w$ or is adjacent to $w$. Hence all such neighbors lie in the
same component of $J'-x$, namely the one containing $w$. Adding $v$
back cannot join this component to any other component. Thus $J-x$
would still be disconnected, contradicting the $2$-connectivity of
$J$.

Let $r=b(J')$, that is, the number of blocks in $J'$ is $r$. Since $w$ is the only possible cutvertex, the $r$ blocks of $J'$ are joined through $w$. Hence $J'-w$ has exactly $r$
components. Every block of $J'$ is either $K_2$ or a $2$-connected cop-win graph, since $J'$ is cop-win. Therefore, by induction, every block $B$ of $J'$ has at least
$2n_B-3$ edges. Thus,
\[|E(J')|
\ge \sum_{B\in\mathcal B(J')}(2n_B-3)
=2\sum_{B\in\mathcal B(J')}(n_B-1)-r
=2(s-2)-r.\]
where the last equality follows from the block identity
\[\sum_{B\in\mathcal B(J')}(n_B-1)=(s-1) -1 = s-2.\]

As $J$ is $2$-connected, $J-w$ is connected. Thus $v$ must
have a neighbor in each of the $r$ components of $J'-w$. Also,
$vw\in E(J)$ because $v$ is dominated by $w$. Hence $d_J(v)\ge r+1.$ It follows that
\[ |E(J)|
=|E(J')|+d_J(v)
\ge 2(s-2)-r+(r+1)
=2s-3. \]

To prove \eqref{copfill_density_con}, let $J$ be any connected cop-win graph. Every block of $J$
is also cop-win, so applying the above bound to each block and adding them results
\[|E(J)|
\ge
\sum_{B\in\mathcal B(J)}(2n_B-3)
= 2|V(J)|-b(J)-2.\]
The same bound also holds when $J$ consists of a single vertex.
\end{proof}

\begin{theorem}\label{copfill_lower}
For every nonempty graph $G$,
 $\copfill(G)\ge 2n-m-b-c-1$.
\end{theorem}
\begin{proof}
Let $F$ be any cop-win completion of $G$, and put $J=G+F$. By
Lemmas~\ref{block_monotonicity} and~\ref{copfill_density},
\[m+|F|
=|E(J)|
\ge 2n-b(J)-2
\ge 2n-(b+c-1)-2.\]
Rearranging gives,
$|F|\ge 2n-m-b-c-1,$ yielding the bound we needed.

Using~\eqref{block_identities} we can get the following equivalent block-wise form,
\begin{equation}
    \copfill(G) \ge 2n - m - b- c- 1 = c-1+\sum_{B\in\mathcal B(G)}(2n_B-3-m_B)
\end{equation}\end{proof}

\subsection{Partial \texorpdfstring{$2$}{2}-trees}\label{section_copfill-partial-two}
A \emph{partial $2$-tree} is a graph of treewidth at most two. A \emph{$2$-tree} starts with $K_2$ and repeatedly adds a vertex adjacent to both endpoints of an existing edge. Clearly it has $2s-3$ edges on $s\ge2$ vertices and is chordal. It is easy to check that treewidth does not increase upon taking minors, and that a partial $2$-tree has a vertex of degree at most two.

For completeness, a tree decomposition consists of vertex bags arranged
on a tree. Every vertex and edge must appear in some bag, and the bags
containing any fixed vertex must form a connected subtree. Its width is
the largest bag size minus one. In a width-two decomposition, after
removing any bag contained in a neighboring bag, a leaf bag contains a
vertex that appears nowhere else. Since the bag has at most three
vertices, this vertex has at most two neighbors. Thus every nonempty
partial $2$-tree has a vertex of degree at most two.

\begin{lemma}\label{partial_2_lemma}
Let $Q$ be a $2$-connected partial $2$-tree with $|V(Q)|\ge3$. Then $Q$ has a degree-two vertex. If $v$ is any such vertex with neighbors $x,y$, then $Q'=(Q-v)+xy$
has treewidth at most two and is either $K_2$ or $2$-connected.
\end{lemma}
\begin{proof}
Because $Q$ is $2$-connected, its minimum degree is at least two. Because it is a partial $2$-tree, it has a vertex of degree at most two. Thus a degree-two vertex $v$ exists. Contracting $vx$ and removing a possible duplicate edge gives exactly $(Q-v)+xy$, so $Q'$ has treewidth at most two.

If $|V(Q')|\ge3$, delete an arbitrary vertex $z$. When $z\notin\{x,y\}$, every path in $Q-z$ that uses $x,v,y$ can instead use the edge $xy$, so $Q'-z$ is connected. When $z=x$ or $z=y$, the vertex $v$ is a leaf of the connected graph $Q-z$. Deleting a leaf from a connected graph does not disrupt connectivity. Therefore, $Q'-z$ is connected. Hence $Q'$ is $2$-connected. If only two vertices remain, that is $|V(Q')| =2$, the added or existing edge $xy$ makes $Q'=K_2$.
\end{proof}

\begin{proposition}\label{partial_2_prop}
Let $B$ be a $2$-connected partial $2$-tree on $s\ge3$ vertices. Repeatedly applying Lemma~\ref{partial_2_lemma}, recording every missing edge $xy$ that is added, produces a set $F_B$ such that $B+F_B$ is a $2$-tree and $|F_B|=2s-3-|E(B)|.$

Therefore, $F_B$ is a minimum cop-win completion of $B$.
\end{proposition}
\begin{proof}
Repeated application of Lemma~\ref{partial_2_lemma} finally results in a $K_2$. When we reverse the deletion process, each deleted vertex is added back next to
the two endpoints of an edge that is already present. This is exactly
a valid step in the construction of a $2$-tree.

Also, any edge added during the forward process joins vertices that had
not yet been deleted at that time. Hence no added edge is lost when the
deleted vertices are restored. Therefore the reverse process gives
exactly $B+F_B$.

A $2$-tree on $s$ vertices has exactly $2s-3$ edges. Since the final
graph is $B+F_B$, we therefore get the above bound.
To see that this completion is minimum, let $F$ be any cop-win
completion of $B$. Since adding edges cannot destroy $2$-connectivity,
$B+F$ is still $2$-connected. By Lemma~\ref{copfill_density},
$|E(B)|+|F|\ge 2s-3.$
Thus $|F|\ge 2s-3-|E(B)|=|F_B|,$
so $F_B$ is a minimum cop-win completion.
\end{proof}

Lemma~\ref{partial_2_lemma} shows that we can repeatedly remove a
degree-two vertex after adding an edge between its two neighbors, until
only $K_2$ remains. Proposition~\ref{partial_2_prop} shows that this
process gives a minimum completion and the exact number of added edges.
The pseudocode below in Algorithm~\ref{partial_2_algo} implements this process.

\begin{algorithm}[!ht]
\caption{Minimum cop-win completion of a partial $2$-tree block}
\label{partial_2_algo}
\begin{algorithmic}[1]
\Statex \textbf{Input:} A $2$-connected partial $2$-tree $B$ with $|V(B)|\ge3$.
\Statex \textbf{Output:} A minimum fill set $F_B$, with $B+F_B$ a $2$-tree.
\State $Q\gets B$; $F_B\gets\varnothing$
\While{$|V(Q)|>2$}
 \State Choose a degree-two vertex $v$ with neighbors $x,y$.
 \If{$xy\notin E(Q)$}
  \State Add $xy$ to $Q$ and to $F_B$.
 \EndIf
 \State Delete $v$ from $Q$.
\EndWhile
\State \Return $F_B$
\end{algorithmic}
\end{algorithm}

Correctness and optimality follow from Proposition~\ref{partial_2_prop}.
In particular, the algorithm adds exactly $2s-3-|E(B)|$ edges.

\begin{theorem}\label{partial_2_theorem}
For every nonempty partial $2$-tree $G$,
\begin{equation}\label{partial_2_eqn}
 \copfill(G)=2n-m-b-c-1
 =c-1+\sum_{B\in\mathcal B(G)}(2n_B-3-m_B).
\end{equation}
The optimum value is computable in $O(n+m)$ time on this class. An optimum completion can be constructed in $O(n^2)$ time and space, and the completed graph is chordal.
\end{theorem}
\begin{proof}
Apply Proposition~\ref{partial_2_prop} to each block of order at least three and leave bridge blocks unchanged. Join one representative from each original component in a path, using $c-1$ new edges. These edges are bridges, and the completed original blocks are $2$-trees. Every cycle lies in a completed block, so the resulting graph is connected and chordal, therefore cop-win. The number of added edges is exactly the right-hand side of~\eqref{partial_2_eqn}, which matches Theorem~\ref{copfill_lower}.

The connected components and blocks of $G$ can be found in $O(n+m)$
time. This gives $b$ and $c$, while $n$ and $m$ are obtained directly
from the input. Hence the formula
$\copfill(G)=2n-m-b-c-1$
gives the optimum value in linear time. 

An adjacency-matrix implementation takes $O(n_B^2)$ time and space per block. Since
\[
 \sum_B n_B=n-c+b\le2(n-c)
 \implies \sum_B n_B^2\le\bigl(\sum_B n_B\bigr)^2=O(n^2),
\]
the total construction has the claimed bounds.

For a block with $s$ vertices, use an adjacency matrix and store
which vertices are still active and their current degrees. In each
iteration, scan for an active degree-two vertex $v$ and find its two
neighbors $x,y$. This takes $O(s)$ time. Checking or adding the edge
$xy$ and updating the degrees takes $O(1)$ time. Since there are
$s-2$ iterations, the algorithm takes $O(s^2)$ time and space.

Apply the algorithm separately to every nontrivial block, and then join
the original connected components by a path. As shown above, summing over all blocks gives
$O(n^2)$ total time and space and $O(n)$ added edges.
\end{proof}

\paragraph*{Recognition on arbitrary inputs}
The same procedure can also test whether a graph is a partial $2$-tree.
Process each $2$-connected block separately. If a current block has more
than two vertices but no degree-two vertex, reject.

If the procedure finishes successfully, reverse the deletion sequence.
Start with the final edge $xy$. Whenever a deleted vertex $v$ is restored,
attach the bag $\{v,x,y\}$ to a bag already containing $x$ and $y$.
This gives a tree decomposition of width at most two, so the original
graph is a partial $2$-tree.

Using the adjacency-matrix implementation above, this recognition takes
$O(n^2)$ time. The $O(n+m)$ bound in
Theorem~\ref{partial_2_theorem} applies when the input is already known
to be a partial $2$-tree.

\subsection{Outerplanar Graphs}\label{section_copfill-outerplanar}
For outerplanar graphs, Theorem~\ref{partial_2_theorem} already gives the optimum value, since outerplanar graphs are partial $2$-trees. Using an outerplanar embedding, we can construct an optimum completion in linear time instead of quadratic and keep the completed graph outerplanar.

\begin{theorem}\label{thm_copfill-outer}
For every nonempty outerplanar graph, an optimum cop-win completion can be constructed in $O(n+m)$ time and space, with the completed graph itself outerplanar.
\end{theorem}
\begin{proof}
Fix an outerplanar embedding. For each block of order at least three, add diagonals to fan-triangulate every original bounded face. These diagonals do not cross, and the completed block is chordal. Using Euler's formula on $B$ we get that this uses
\[\sum_{f\text{ bounded}}(|f|-3)
=(2m_B-n_B)-3(m_B-n_B+1)=2n_B-3-m_B.\]
edges in block $B$. Join the original components by $c-1$ bridges. The number of added edges matches Theorem~\ref{partial_2_theorem}, so it is optimal. The block decomposition, embedding, and face traversal take linear time. The implementation is given below.
\end{proof}

Algorithm~\ref{outerplanar_algo} works separately on the bounded
faces of each block in an outerplanar embedding. Each face is triangulated as a fan
using only diagonals inside that face. This avoids problems caused by bridges
or by existing diagonals in other parts of the block.

\begin{algorithm}[!ht]
\caption{Minimum cop-win completion of an outerplanar graph}
\label{outerplanar_algo}
\begin{algorithmic}[1]
\Statex \textbf{Input:} A nonempty outerplanar graph $G$.
\Statex \textbf{Output:} A minimum fill set $F$, preserving outerplanarity.
\State Compute components, blocks, and an outerplanar embedding.
\State $F\gets\varnothing$
\For{each block $B$ with $|V(B)|\ge3$}
 \For{each original bounded face of the induced embedding of $B$}
  \State List its boundary in cyclic order as $v_1,\ldots,v_\ell$.
  \For{$j=3,\ldots,\ell-1$}
   \State Add $v_1v_j$ to $F$.
  \EndFor
 \EndFor
\EndFor
\State Choose a vertex $r_i$ of every original component, $1\le i\le c$.
\For{$i=1,\ldots,c-1$}
 \State Add $r_ir_{i+1}$ to $F$.
\EndFor
\State \Return $F$
\end{algorithmic}
\end{algorithm}

Every added diagonal is new, since an existing edge between two
nonconsecutive vertices would already split the face. Diagonals added in different faces are distinct and do not cross.

The total length of all face boundaries is $O(m)$, so all faces can be processed and all new edges can be written in $O(n+m)$ time using
adjacency lists and an outerplanar embedding. Finally, place the connected components side by side and join them by bridges drawn outside the components. This preserves outerplanarity.

Hence the construction in Theorem~\ref{thm_copfill-outer} takes $O(n+m)$ time and space.

\section{A triangle-space bound}\label{section_copfill-block-cactus}
The lower bound from Theorem~\ref{copfill_lower} can become weak when the graph contains large clique blocks, even though such blocks
need no added edges. To handle graphs containing both dense blocks and long cycles, we use a second lower bound based on cycles that are not
generated by triangles.

\subsection{A Lower Bound from the Cycle Space}\label{subsection_copfill-cycle-space}
We view each edge set as a vector over $\mathbb F_2$. The cycle space
$\mathcal C(G)$ is the set of edge sets in which every vertex has even
degree. Its dimension is $m-n+c$,
which follows from the usual fundamental-cycle basis obtained from a spanning forest. Let $\mathcal T(G)$ be the span of all triangle vectors, and put $\delta(G)=\dim\mathcal C(G)-\dim\mathcal T(G).$

\begin{lemma}\label{copfill_triangle_lemma}
If $J$ is cop-win, then $\mathcal C(J)=\mathcal T(J)$.
\end{lemma}
\begin{proof}
We use induction on a dismantling order of $J$. Let $v$ be dominated by
$w$, and take any $z\in\mathcal C(J)$.
For every edge $vu$ of $z$ with $u\ne w$, add the triangle $vuw$ to
$z$. This triangle exists because $w$ dominates $v$, so every neighbor
$u$ of $v$ is also adjacent to $w$. After doing this, all edges of $z$
incident with $v$, except possibly $vw$, are removed.

Since $z$ has even degree at $v$, the edge $vw$ must also disappear.
Thus the remaining edge set can be considered as a cycle-space element of $J-v$.

By induction, this remaining element is a sum of triangles in $J-v$.
Adding back the triangles used above shows that the original $z$ is also
a sum of triangles in $J$.
\end{proof}
Let $E_0(G)$ be the set of edges lying in no triangle of $G$. For $z\in\mathcal C(G)$, define $L_0(z)=|\supp(z)\cap E_0(G)|$. Note that every vector of $\mathcal T(G)$ vanishes on $E_0(G)$.
\begin{theorem}\label{triangle_space_bound}
For every nonempty graph $G$ and every $z\in\mathcal C(G)$,
\begin{equation}
 \copfill(G)\ge c(G)-1+L_0(z)-3\delta(G).
\end{equation}
\end{theorem}
\begin{proof}
Let $F$ be a cop-win completion of size $k$, and put $J=G+F$.
We view every vector of $\mathcal C(G)$ as a vector of $\mathcal C(J)$
by giving all coordinates corresponding to new edges the value zero.

Choose a basis of $\mathcal T(G)$ consisting of triangles of $G$.
Since $J$ is cop-win, Lemma~\ref{copfill_triangle_lemma} gives
$\mathcal C(J)=\mathcal T(J)$. Hence we can extend this basis to a
basis of $\mathcal C(J)$ by adding triangles of $J$. Taking $\delta=\delta(G)$, we get that the number of triangles that have to be added is
\[\dim\mathcal C(J)-\dim\mathcal T(G)
=(m+k-n+1)-(m-n+c-\delta) 
=k-c+1+\delta.\]

Now write $z$ as a sum of elements of this basis. Let
$T_1,\ldots,T_q$ be the newly added triangles that occur in this sum, and let $U=\operatorname{span}\{T_1,\ldots,T_q\}.$
Since these triangles are part of a basis,
\[
q\le k-c+1+\delta,\qquad
\dim U=q,\qquad
U\cap\mathcal T(G)=\{0\}.
\]
We next look only at the new edges in $F$. Define
$\rho:U\longrightarrow \mathbb F_2^F$,
by keeping only the entries corresponding to edges of $F$.
If a vector of $U$ is in $\ker\rho$, then it uses no edge of $F$ and
therefore belongs to $\mathcal C(G)$. Therefore, $\ker\rho \subseteq \mathcal C(G)$. Also, $\ker\rho\cap\mathcal T(G)=\{0\}$. Therefore,
$\dim\ker\rho+\dim\mathcal T(G)
\le \dim\mathcal C(G).$

Recall that $\delta=\dim\mathcal C(G)-\dim\mathcal T(G),$ therefore from the above inequality
we get $\dim\ker\rho\le\delta$. Thus by rank-nullity theorem,
$\operatorname{rank}\rho
=\dim U-\dim\ker\rho
\ge q-\delta.$

Let $a_F$ be the number of different fill edges that occur in
$T_1,\ldots,T_q$. Since $\rho(U)$ uses only these $a_F$ coordinates (the remaining $F$ coordinates are zero),
$a_F\ge\operatorname{rank}\rho\ge q-\delta.$

The expansion of $z$ can be written as,
$T_1+\cdots+T_q=z+t_0$, for some $t_0\in\mathcal T(G)$.
Since neither $z$ nor $t_0$ contains a fill edge, every fill edge
appearing among $T_1,\ldots,T_q$ must appear an even number of times.
Thus each of the $a_F$ such edges appears at least twice.

Now consider an edge in $\supp(z)\cap E_0(G)$. Such an edge appears in
$z$, but it cannot appear in $t_0$, because no triangle of $G$ contains
an edge of $E_0(G)$. Hence it must appear at least once among
$T_1,\ldots,T_q$.

The $q$ triangles contain $3q$ edge occurrences in total, and the $L_0(z)$ original edges of $G$ are disjoint from the $a_F$ fill edges. Therefore
\[3q\ge L_0(z)+2a_F
   \ge L_0(z)+2(q-\delta).\]
Hence, $L_0(z)\le q+2\delta\le(k-c+1+\delta)+2\delta=k-c+1+3\delta$.
Rearranging this gives, $k\ge c-1+L_0(z)-3\delta,$
which proves the result.
\end{proof}

\subsection{Block-cactus Graphs}\label{subsection_copfill-block-cactus-formula}
A \emph{block-cactus graph} has every block a clique or a chordless cycle~\cite{rautenbach1998domatic}. Bridge blocks and triangles are treated as clique blocks. Let $C_1,\ldots,C_r$ be its cycle blocks of lengths $\ell_i\ge4$, and set $L=\sum_i\ell_i$.

\begin{theorem}\label{copfill_block_cactus}
For every nonempty block-cactus graph $G$,
\begin{equation}
 \copfill(G)=c-1+\sum_{i=1}^r(\ell_i-3)=c-1+L-3r.
\end{equation}
An optimum completion can be constructed in $O(n+m)$ time and space and is chordal.
\end{theorem}
\begin{proof}
Every cycle of $G$ is contained in a single block, so the cycle space of $G$ is obtained by combining the cycle spaces of its blocks. For a clique block, every cycle can be written as a sum of triangles, so it contributes nothing to $\delta(G)$. A bridge contains no cycle, so it also contributes nothing. Each long cycle block $C_i$, however, has a one dimensional cycle space and contains no triangle. Therefore, each $C_i$ contributes exactly one to $\delta(G)$. Since there are $r$ such cycle blocks, we get $\delta(G)=r.$

Take $z$ to be the union of the edge sets of $C_1,\ldots,C_r$. They are edge-disjoint even subgraphs, so $z\in\mathcal C(G)$. No edge of these blocks lies in a triangle, giving $L_0(z)=L$. Theorem~\ref{triangle_space_bound} now gives $\copfill(G)\ge c-1+L-3r$.

For equality, fan-triangulate each $C_i$ by adding $\ell_i-3$ diagonals (as we did for the outerplanar case), leave the clique blocks unchanged, and connect the original components using $c-1$ bridges. The resulting graph is connected and chordal, and the number of added edges matches the lower bound. The block decomposition, cycle traversal, and construction of the completion can all be done in linear time since $L\le \sum_B n_B\le 2(n-c).$

The recognition test and the full algorithm are given below.
\end{proof}

\paragraph*{Block-cactus completion and recognition}
A block is a clique exactly when
$m_B=\binom{n_B}{2}.$
If it is not a clique, then it is a chordless cycle exactly when every
vertex has degree two inside the block. Thus, after computing the block
decomposition, we can recognize block-cactus graphs in linear time by
checking these two cases for every block. If some block is neither a
clique nor a chordless cycle, we reject. Otherwise, we construct an
optimum completion using Algorithm~\ref{copfill-block-cactus_algo}.

\begin{algorithm}[!ht]
\caption{Minimum cop-win completion of a block-cactus graph}
\label{copfill-block-cactus_algo}
\begin{algorithmic}[1]
\Statex \textbf{Input:} A nonempty block-cactus graph $G$.
\Statex \textbf{Output:} A minimum fill set $F$, producing a chordal graph.
\State Compute components and blocks; set $F\gets\varnothing$.
\For{each block $B$}
 \If{$m_B\ne\binom{n_B}{2}$}
  \State List its cycle in order as $v_1,\ldots,v_\ell$.
  \For{$j=3,\ldots,\ell-1$}
   \State Add $v_1v_j$ to $F$.
  \EndFor
 \EndIf
\EndFor
\State Choose a vertex $r_i$ of every original component, $1\le i\le c$.
\For{$i=1,\ldots,c-1$}
 \State Add $r_ir_{i+1}$ to $F$.
\EndFor
\State \Return $F$
\end{algorithmic}
\end{algorithm}
Theorem~\ref{copfill_block_cactus} proves that the constructed completion is
optimal even when added edges are allowed anywhere in the graph. Traversing
all cycle blocks takes $O(L)$ time, where
\[
L\le \sum_B n_B\le 2(n-c)=O(n).
\]
The blocks can be found and the clique blocks checked in $O(n+m)$ time.
Since only $O(n)$ edges are added, the whole construction uses
$O(n+m)$ time and space.

\paragraph*{Why the triangle-space bound is useful}
For a cycle block of length $\ell$, the density term is $\ell-3$. For a clique block of order $s$, it is
\[
 2s-3-\tbinom s2=-\frac{(s-2)(s-3)}2.
\]
Thus, a large clique can make the density bound much weaker, even though the clique itself needs no added edges. In contrast, a clique contributes nothing to $\delta(G)$.

For example, consider a $4$-cycle and a clique $K_s$, where $s\ge4$, sharing exactly one vertex. The triangle-space bound gives the exact value $1$, while the density bound gives
$1-\frac{(s-2)(s-3)}{2}.$

Hence, the triangle-space bound captures the contribution of the cycle without being weakened by the presence of a large clique.

\begin{corollary}\label{cor_chor-opt}
If $G$ is a nonempty partial $2$-tree or block-cactus graph, then
\[
 \copfill(G)=\fillin(G)+c(G)-1.
\]
Therefore, the two optima coincide on connected graphs in these classes.
\end{corollary}
\begin{proof}
Theorems~\ref{partial_2_theorem} and~\ref{copfill_block_cactus} produce a connected chordal completion of size $\copfill(G)$. The minimum size of such a completion is $\fillin(G)+c(G)-1$, as observed in Section~\ref{section_prelim}. The reverse inequality holds for every graph, since every connected chordal graph is cop-win.
\end{proof}

\section{Disjoint unions, joins, and cographs}\label{section_composition}
The previous results give graph classes for which an optimum chordal completion is also an optimum cop-win completion. We next look at two simple ways of combining graphs. Besides giving more exact algorithms, these formulas show that cop-win
completion can require fewer added edges than chordal completion for some classes of graphs, and the gap between the two optimum values can be arbitrarily large.

\subsection{Disjoint unions}
We have so far joined the connected components by bridges when constructing completions. The following result shows that this is always optimal, provided each component is completed optimally.

We give the details needed for the lower bound in Theorem~\ref{union_theorem}. The main idea is to contract vertices along a dismantling order while keeping track of edges between the two sides.

\begin{lemma}\label{contraction_lemma}
Let $J$ have a dismantling order $v_1,\ldots,v_s$. Suppose $z=v_i$ is dominated at its deletion by a later vertex $w=v_j$, where $i<j$. Contracting $zw$ gives another cop-win graph, and the original order with $z$ omitted is a dismantling order of the contracted graph.
\end{lemma}
\begin{proof}
Follow the original order before $z$. At every such step, both $z$ and $w$ are still present. If a vertex $v$ was dominated by $u\notin\{z,w\}$, this remains true after contraction. Indeed, the new neighborhood of each vertex is the image of its old neighborhood, and neighborhood inclusion is preserved. If $v$ was dominated by $z$ or $w$, the merged vertex dominates it, because its neighborhood contains the images of both neighborhoods.

When the order reaches $z$, all its remaining neighbors are already neighbors of $w$, since $w$ dominates $z$. Thus contracting $z$ into $w$ has the same effect on the remaining vertices as deleting $z$. We can therefore continue with the rest of the original dismantling order.
\end{proof}

\begin{lemma}\label{cut_lemma}
Let $J$ be a cop-win graph and let $V(J)=A\sqcup B$, where $A$ and $B$ are nonempty. There are cop-win supergraphs $J_A$ of $J[A]$ and $J_B$ of $J[B]$ such that
\[ |E(J_A)\setminus E(J[A])|+|E(J_B)\setminus E(J[B])|
 \le |E_J(A,B)|-1.\]
\end{lemma}
\begin{proof}
Choose a dismantling order of $J$, and for every vertex other than the last choose a later vertex $p(v)$ that dominates it when it is deleted. The edges $vp(v)$ form a tree $T$ rooted at the last vertex. Without loss of generality assume that the root is in $B$.

We will first define two maps. The map $q_B$ fixes the vertices of $B$ and sends each vertex of $A$ to its first ancestor in $B$ in the tree $T$. Let $T_0$ be the component of $T[B]$ containing the root. Since $A$ is nonempty, there is a vertex $a_0\in A$ whose parent belongs to $T_0$. Define $q_A$ to fix $A$, send every vertex of $T_0$ to $a_0$, and send every other vertex of $B$ to its first ancestor in $A$.

The vertices mapped to any fixed vertex by either map form a connected subtree of $T$. We can therefore obtain the quotient graph for each map by contracting parent edges within these subtrees. Lemma~\ref{contraction_lemma} shows that both quotient graphs are cop-win. Write them as $J_A$ and $J_B$. Since the maps fix the vertices of their target sets, $J_A$ contains $J[A]$ and $J_B$ contains $J[B]$.

We now count the new edges of the two quotient graphs. Fix one of the target sets $C\in\{A,B\}$ and its map $q_C$. Consider an edge of the quotient that was not present in $J[C]$. It is the image of an edge $xy\in E(J)$, with $q_C(x)\ne q_C(y)$. Choose such an edge and call its earlier endpoint $x$ and its later endpoint $y$ in the dismantling order.

If $x\notin C$, replace $x$ by its parent $p(x)$. The edge to $y$ is still present, because $p(x)$ dominates $x$ when $x$ is deleted and $y$ is later than $x$. Also $q_C(p(x))=q_C(x)$ whenever $x\notin C$, so its image does not change. If the order of the two endpoints changes, rename the earlier one $x$. Repeat this until the earlier endpoint belongs to $C$. The process terminates since an endpoint moves strictly later in the dismantling order at every step. It cannot end with both endpoints in $C$, since the image would then be an original edge of $J[C]$. We therefore end with an edge across $(A,B)$ whose earlier endpoint belongs to $C$.

We associate each missing quotient edge with the cross edge found above. For a fixed target set, different missing quotient edges give different cross edges, since each cross edge has only one image under $q_C$. Moreover, the cross edges used for $J_A$ and $J_B$ are distinct, since the earlier endpoint lies in $A$ in the first case and in $B$ in the second.

Finally, the tree edge $a_0p(a_0)$ is a cross edge whose endpoints are mapped to the same vertex under both maps (to $a_0$ by $q_A$ and to $p(a_0)$ by $q_B$). Hence it cannot correspond to a missing edge in either quotient graph. Therefore, the total number of missing edges in $J_A$ and $J_B$ is at most $|E_J(A,B)|-1$, giving the required bound.
\end{proof}

\begin{theorem}\label{union_theorem}
Let $G_1,\ldots,G_q$ be nonempty graphs on pairwise disjoint vertex sets. Then
\[ \copfill\left(\bigsqcup_{i=1}^q G_i\right)
 =\sum_{i=1}^q\copfill(G_i)+q-1.\]
An optimum completion is obtained by completing each $G_i$ optimally and joining the last vertices of their chosen dismantling orders by $q-1$ bridges forming a tree.
\end{theorem}
\begin{proof}
The above construction gives the upper bound. Complete each $G_i$ optimally
and choose a last vertex $r_i$ of a dismantling order of the completed
graph. Join the vertices $r_1,\ldots,r_q$ by $q-1$ bridges forming a tree.
We can first dismantle each completed $G_i$ to $r_i$. The remaining graph
is a tree on the vertices $r_1,\ldots,r_q$, which is cop-win. Hence the
whole completed graph is cop-win.
For the lower bound, it is enough to consider two graphs. Let $F$ be any
cop-win completion of $G_1\sqcup G_2$, put
$J=(G_1\sqcup G_2)+F,$ and let $h$ be the number of edges of $F$ joining the two vertex sets.

We use Lemma~\ref{cut_lemma}, proved above. It gives cop-win supergraphs of $J[V(G_1)]$ and $J[V(G_2)]$ using together at most $h-1$ additional edges. All edges already added inside the two original graphs, together with these further edges, therefore give separate cop-win completions using at most $|F|-1$ edges in total. Hence
$\copfill(G_1)+\copfill(G_2)\le |F|-1.$

Applying this repeatedly proves the result for $q$ graphs.
\end{proof}

Therefore, note that removing an isolated vertex, or a clique component, decreases $\copfill$ by exactly one whenever another component remains. We will use this when handling twin cover.

\subsection{Joins and cographs}
For graphs $H_1,\ldots,H_q$ on disjoint vertex sets, their \emph{join} $H_1\vee\cdots\vee H_q$ is obtained by adding all edges between different graphs. Cameron and Fitzpatrick~\cite{cameron2015edge} showed that a join of two graphs is cop-win exactly when at least one of them is cop-win. The same statement holds for any number of factors.

\begin{proposition}\label{join_prop}
For nonempty graphs $H_1,\ldots,H_q$, where $q\ge2$,
\[
 \copfill(H_1\vee\cdots\vee H_q)=\min_{1\le i\le q}\copfill(H_i).
\]
\end{proposition}
\begin{proof}
All edges between different factors are already present, so every fill edge lies inside one factor. If the completed join is cop-win, the result of Cameron and Fitzpatrick implies that some completed factor is cop-win. Thus every completion needs at least $\min_i\copfill(H_i)$ edges. Conversely, complete only a factor having the smallest optimum. The join then becomes cop-win and uses exactly that many edges.
\end{proof}

A \emph{cograph} can be obtained from single vertices by repeatedly taking disjoint unions and joins. Therefore, the formulas from Theorem~\ref{union_theorem} and Proposition~\ref{join_prop} can be used to get an algorithm for the whole class.

\begin{theorem}\label{cograph_thm}
The optimum cop-win completion of a cograph can be found in polynomial time, and an optimum set of added edges can also be constructed. Given a cotree, the optimum value can be computed in time linear in the size of the cotree.
\end{theorem}
\noindent
The union and join formulas give the cotree algorithm. Spiders are cop-win, extending the
algorithm to $P_4$-sparse graphs (Corollary~\ref{p4sparse_cor}).
\begin{proof}
Consider a cotree of $G$. Each node of the cotree represents the cograph
obtained from the leaves below it. For every node $x$, let $f(x)$ denote
the minimum number of edges needed to make this graph cop-win. We compute
these values bottom-up.

If $x$ is a leaf, then it represents a single vertex, so $f(x)=0$.

Suppose $x$ is a union node with children $x_1,\ldots,x_q$. The graph
represented by $x$ is the disjoint union of the graphs represented by its
children. Hence, by Theorem~\ref{union_theorem},
\[f(x)=\sum_{i=1}^q f(x_i)+q-1.\]

If $x$ is a join node, then Proposition~\ref{join_prop} gives
\[f(x)=\min_{1\le i\le q} f(x_i).\]

Thus the optimum value at each node is determined directly from the values
of its children, so all values can be computed in one bottom-up traversal
of the cotree.

The completion itself can be constructed at the same time. At a union
node, take optimum completions of all children and join the completed
graphs by $q-1$ bridges. At a join node, choose a child with minimum value,
use an optimum completion only inside that child, and add no edges inside
the other children. Theorem~\ref{union_theorem} and
Proposition~\ref{join_prop} show that these choices are optimal at every
node.

Since a cotree of a cograph can be found in polynomial time, this gives
the claimed polynomial-time algorithm.
\end{proof}

For completeness, Algorithm~\ref{cograph_algo} gives the explicit construction used in Theorem~\ref{cograph_thm}. For each node of the cotree, we store an optimum fill set for the graph represented by that node, its size, and a vertex that can remain last in a dismantling order of the completed graph. This last vertex is used when completed child graphs have to be joined by bridges at a union node.

\begin{algorithm}[!ht]
\caption{Minimum cop-win completion of a cograph from its cotree}
\label{cograph_algo}
\begin{algorithmic}[1]
\Statex \textbf{Input:} A cotree of a nonempty cograph $G$.
\Statex \textbf{Output:} A minimum fill set $F$.
\For{each node $x$ of the cotree in bottom-up order}
    \If{$x$ is a leaf representing a vertex $v$}
        \State $f_x\gets0$; $F_x\gets\varnothing$; $r_x\gets v$
    \ElsIf{$x$ is a union node with children $x_1,\ldots,x_q$}
        \State $F_x\gets\bigcup_i F_{x_i}$
        \For{$i=1,\ldots,q-1$}
            \State Add $r_{x_i}r_{x_{i+1}}$ to $F_x$
        \EndFor
        \State $f_x\gets\sum_i f_{x_i}+q-1$
        \State $r_x\gets r_{x_1}$
    \Else
        \State Choose a child $x_j$ with minimum value $f_{x_j}$
        \State $F_x\gets F_{x_j}$; $f_x\gets f_{x_j}$; $r_x\gets r_{x_j}$
    \EndIf
\EndFor
\State Let $r$ be the root of the cotree.
\State \Return $F_r$
\end{algorithmic}
\end{algorithm}

At a union node, first dismantle each completed child to its stored
representative. The added bridges then leave a path on these
representatives, which can be dismantled to the first one. Hence the
representative stored for the union node is valid.

At a join node, dismantle the chosen minimum-cost child to its stored
representative. Because this vertex is adjacent to every vertex in the
other children, all remaining vertices can then be deleted, so this
representative is also valid.

Therefore the construction preserves a valid last vertex at every node.
Theorem~\ref{union_theorem} and Proposition~\ref{join_prop} show that the
number of added edges is minimum in both cases.

\subsection{Extension to \texorpdfstring{$P_4$}{P4}-sparse graphs}\label{subsection_p4sparse}
\begin{corollary}\label{p4sparse_cor}
An optimum cop-win completion of a $P_4$-sparse graph can be constructed in $O(n+m)$ time.
\end{corollary}
\begin{proof}
A graph is \emph{$P_4$-sparse} if every set of five vertices induces
at most one induced $P_4$. Every nontrivial $P_4$-sparse graph is a
disjoint union, a join, or a spider~\cite{jamison1992tree}.
A spider has a partition $(S,K,R)$ with $|S|=|K|\ge2$, where
$S$ is independent, $K$ is a clique, $R$ is complete to $K$ and
anticomplete to $S$, and the edges between $S$ and $K$ form either
a perfect matching or the complement of a perfect matching in the
complete bipartite graph with parts $S$ and $K$.

Every spider is cop-win. For each $s\in S$, choose a neighbor $v\in K$.
Since $N(s)\subseteq K$ and $K$ is a clique, $N[s]\subseteq N[v]$.
Delete all vertices of $S$. Any vertex of $K$ is then universal in
the remaining graph, which dismantles to that vertex. No completion
of the head $G[R]$ is needed.

Process the decomposition from the leaves to the root. A leaf or
spider has value zero. At a union node with $q$ children,
Theorem~\ref{union_theorem} gives $f=\sum_i f_i+q-1$ and at a join node,
Proposition~\ref{join_prop} gives $f=\min_i f_i$. These recurrences
therefore compute the optimum.

For construction, retain at each node a vertex at which a dismantling
order of its chosen completion can end. At a leaf, retain its vertex and at a spider, retain any vertex of $K$. At union and join nodes, use
the construction from Algorithm~\ref{cograph_algo}. Connect the retained vertices of all children by a path at a union node, and keep only a minimum-value child's completion at a join node. The same dismantling argument proves that these choices preserve a valid retained vertex.

The decomposition has $O(n)$ nodes and can be constructed in $O(n+m)$ time~\cite{jamison1992recognizing}. Store only values, retained vertices, and choices in the bottom-up pass. Recover the fill edges in a second traversal, visiting all children at a union, only the chosen child at a join, and stopping at a leaf or spider. This avoids copying fill sets at intermediate nodes. The total number of child links processed and bridges output is $O(n)$, giving the stated $O(n+m)$ running time.
\end{proof}

\subsection{A gap with chordal completion}
For partial $2$-trees and block-cactus graphs, Corollary~\ref{cor_chor-opt} showed that an optimum connected chordal completion is also optimum for cop-win completion. The equality does not hold for all graphs. Complete bipartite graphs give a simple example, and their values follow from the join formula.

\begin{corollary}\label{bipartite_gap}
For $2\le a\le b$,
\[
 \copfill(K_{a,b})=a-1,
 \qquad
 \fillin(K_{a,b})=\binom{a}{2}.
\]
In particular, the gap between the two optima can be arbitrarily large.
\end{corollary}
\begin{proof}
Since $K_{a,b}=\overline K_a\vee\overline K_b$ and an independent set on $s$ vertices needs $s-1$ edges to become cop-win, Proposition~\ref{join_prop} gives $\copfill(K_{a,b})=a-1$. An optimum completion adds a spanning tree inside the smaller part.

For a chordal completion, at least one part must be made a clique. Otherwise, two nonadjacent vertices from each part would induce a $4$-cycle. Making the smaller part a clique uses $\binom a2$ edges and is sufficient, proving the second formula.
\end{proof}

This also shows that the equality for partial $2$-trees does not extend
to treewidth three graphs. $\operatorname{tw}(K_{3,3})=3$, but
$\copfill(K_{3,3})=2$ whereas $\fillin(K_{3,3})=3$.
For $K_{a,a}$, the ratio between the two optima is $a/2$, and hence is
unbounded. More generally, the join formula gives
$\copfill(K_{a_1,\ldots,a_q})=\min_i(a_i-1)$ for complete multipartite
graphs with $q\ge2$.

\section{Parameterized algorithms under structural restrictions}\label{section_direct-fpt}
The hardness reduction shows that a small dominating set does not make the problem tractable, even when the completion budget is also a parameter. We now consider parameters that let us restrict where edges may be added. We first use vertices with the same neighbors, and then extend the idea to twin cover. Finally, we give an algorithm using the completion budget and treewidth together.

\subsection{Neighborhood diversity}
The notion of neighborhood diversity was introduced by Lampis~\cite{lampis2012algorithmic}. Two distinct vertices $u$ and $v$ have the same type if they have the same neighbors apart from each other, that is,
$N_G(u)\setminus\{v\}=N_G(v)\setminus\{u\}$.
This relation partitions $V(G)$ into types. The number of types is the \emph{neighborhood diversity} $d=\nd(G)$. Each type is a clique or an independent set, and any two types are either completely adjacent or have no edges between them. Thus vertices within a type are interchangeable. Comparing all pairs of neighborhoods gives the partition in $O(n^3)$ time.

\begin{lemma}\label{fpt_nd_lemma}
For $k\ge1$, choose $\min\{|V_i|,2k\}$ arbitrary representatives $R_i$ in each type $V_i$, and put $R=\bigcup_iR_i$. If a cop-win completion of size at most $k$ exists, then an optimum completion exists with both endpoints of every added edge in $R$. Moreover, $|R|\le2kd$.
\end{lemma}
\begin{proof}
Let $F$ be a completion with $|F|\le k$, and let $T$ contain the endpoints of its edges. Then $|T|\le2k$. For each type $V_i$, we can permute its vertices so that those in $T\cap V_i$ are mapped into $R_i$. These permutations give an automorphism $\pi$ of $G$. Replacing each $uv\in F$ by $\pi(u)\pi(v)$ gives a set of the same size, with endpoints in $R$, and the completed graph is isomorphic to $G+F$. Hence it is also cop-win.
\end{proof}

\begin{theorem}\label{fpt_nd_thm}
For $k\ge1$, \problem{Cop-Win Edge Completion} can be solved in time
$O\bigl((k+1)(2kd)^{2k}n^3\bigr)$, where $d=\nd(G)$. For $k=0$, polynomial-time cop-win recognition suffices. Thus the problem is FPT for $p=k+d$, in time $2^{O(p\log p)}n^{O(1)}$.
\end{theorem}
\begin{proof}
Assume first that $k\ge1$. Compute the representative set $R$ from
Lemma~\ref{fpt_nd_lemma}. By the lemma, if $G$ has a cop-win completion
of size at most $k$, then it has one in which every added edge has both
endpoints in $R$. Hence it is enough to consider only nonedges whose
endpoints lie in $R$.

Let $q_R$ be the number of such nonedges. We try every set $F$ of at
most $k$ of these nonedges and test whether the full graph $G+F$ is
cop-win. If some such set succeeds, we return yes. Conversely, if a
completion of size at most $k$ exists, Lemma~\ref{fpt_nd_lemma} guarantees
one with all endpoints in $R$, so it is among the sets considered.
Thus the search is correct.

Since $q_R\le |R|^2$, the number of sets considered is at most
\[
\sum_{i=0}^{\min\{k,q_R\}}\binom{q_R}{i}
\le (k+1)\max\{1,q_R\}^{k}
\le (k+1)\max\{1,|R|\}^{2k}.
\]
By Lemma~\ref{fpt_nd_lemma}, $|R|\le2kd$, and therefore the number of
choices is at most $(k+1)(2kd)^{2k}.$
For each choice, cop-win recognition takes $O(n^3)$ time, giving a total
running time of $O\bigl((k+1)(2kd)^{2k}n^3\bigr).$

Finally, let $p=k+d$. Since $k,d\le p$,
$(k+1)(2kd)^{2k}
\le (p+1)(2p^2)^{2p}
=2^{O(p\log p)}.$
Hence the problem is FPT parameterized by $p=k+d$.

For $k=0$, no edges may be added, so it is enough to test directly
whether $G$ is already cop-win.
\end{proof}

This argument only uses automorphisms and a polynomial-time test for the target property. It therefore also applies to edge completion into any isomorphism-invariant graph class that can be recognized in polynomial time.

\subsection{Twin cover}\label{subsection_twin-cover}
Neighborhood diversity treats all vertices of one type as interchangeable. Twin cover lets us use a similar idea. The parameter was introduced by Ganian~\cite{ganian2015improving}.

Two adjacent vertices $u,v$ are \emph{true twins} if $N_G[u]=N_G[v]$. A set $X\subseteq V(G)$ is a \emph{twin cover} if every edge with both endpoints outside $X$ joins true twins. The minimum size of such a set is denoted by $\operatorname{tc}(G)$. Every vertex cover is a twin cover, so
$\operatorname{tc}(G)\le\vc(G)$.
For a twin cover $X$, each component of $G-X$ is a clique, and all vertices of that clique have the same neighbors in $X$. We call these the \emph{outside cliques}.

\begin{lemma}\label{tc_truncation_lemma}
Let $X$ be a twin cover of $G$ and let $k\ge1$. From every outside clique $Q$, keep any $\min\{|Q|,2k+1\}$ vertices, and let $G'$ be the resulting induced subgraph, including all of $X$. Then
 \[\copfill(G)\le k\quad\iff\quad\copfill(G')\le k.\]
Moreover, any completion of $G'$ using at most $k$ edges extends to one of $G$ using the same added edges.
\end{lemma}
\begin{proof}
Suppose first that $G$ has a cop-win completion $F$ with $|F|\le k$. At most $2k$ vertices are endpoints of edges in $F$. Since all vertices of an outside clique are true twins, we may permute the clique so that every fill endpoint lies among the retained vertices. In every clique that was shortened, at least one retained vertex $u$ is therefore untouched by $F$. Every removed vertex $v$ has the same original closed neighborhood as $u$ and is also untouched by $F$, so $N[v]\subseteq N[u]$ in the completed graph. Hence all removed vertices can be deleted first. What remains is a cop-win completion of $G'$, proving the forward direction.

Conversely, let $F$ be a cop-win completion of $G'$ with $|F|\le k$, and restore all removed vertices with their original adjacencies. In each shortened outside clique, at least one retained vertex $u$ is not incident with any edge of $F$. Every restored vertex $v$ is a true twin of $u$ in
the original graph and is also untouched by $F$, so $u$ dominates $v$. We can therefore delete all restored vertices first. After these deletions, the remaining graph is exactly $G'+F$, which is cop-win. Continuing with a dismantling order of $G'+F$ gives a dismantling order of $G+F$. Hence $G+F$ is cop-win, so the same fill set $F$ is also a completion of $G$.
\end{proof}

\begin{theorem}\label{fpt_tc_budget_thm}
Given a twin cover $X$ of size $h$, \problem{Cop-Win Edge Completion} can be solved in time
$2^{O(k(h+\log(k+1)))}n^{O(1)}$. An appropriate completion is returned when one exists.
\end{theorem}
\begin{proof}
If $k=0$, we only need to test whether $G$ is already cop-win.
Assume therefore that $k\ge1$.

First apply Lemma~\ref{tc_truncation_lemma}. This gives a graph $G'$
in which every outside clique has size at most $2k+1$, without changing
whether a completion of size at most $k$ exists.

Let $X$ be the twin cover, with $|X|=h$. We group the outside cliques of
$G'$ according to two properties. First, their common neighborhood
$S\subseteq X$ and second, their size $s$. Since $s\le2k+1$, there are at most
$2^h(2k+1)$ such classes. Cliques in the same class have exactly the same
adjacencies to the rest of the graph and can therefore be permuted.

For each class $\mathcal C$, choose $\min\{2k,|\mathcal C|\}$ representative cliques, and let $R$
contain $X$ together with all vertices of these chosen cliques. A fill set
of at most $k$ edges has at most $2k$ endpoints, so it can touch at most
$2k$ outside cliques. Hence, by permuting cliques within each class, we may
assume that every outside clique touched by the fill set is one of the
chosen representatives. Therefore, if a solution exists, then one exists
with every fill-edge endpoint in $R$.

The size of $R$ is bounded by
\[
|R|
\le
h+2k\,2^h\sum_{s=1}^{2k+1}s
=
h+O(k^3 2^h).
\]
We now try every set $F$ of at most $k$ nonedges with both endpoints in
$R$, and for each choice test whether the full graph $G'+F$ is cop-win.

There are at most
\[
(k+1)\max\{1,|R|\}^{2k}
=
2^{O(k(h+\log(k+1)))}
\]
such choices, and each can be tested in polynomial time. Thus the total
running time is
\[
2^{O(k(h+\log(k+1)))}n^{O(1)}.
\]

Finally, if a completion of $G'$ is found, Lemma~\ref{tc_truncation_lemma}
allows the removed vertices to be restored without adding any further
edges. Hence the same fill set gives a completion of the original graph
$G$.
\end{proof}

We can now remove the dependence on $k$. The main point is that, after handling connected components, making a twin cover into a clique already gives a completion whose size depends only on the cover.

\begin{theorem}\label{fpt_tc_thm}
\problem{Cop-Win Edge Completion} is FPT parameterized by just $t=\operatorname{tc}(G)$ and can be solved in
$2^{O(t^3)}n^{O(1)}$ time. The algorithm can also find an optimum completion.
\end{theorem}
\begin{proof}
By Theorem~\ref{union_theorem}, we may solve the connected components separately and then join their completions by bridges. We therefore consider a connected graph $G$. If $G$ is a clique, its optimum is zero.

Otherwise, form an auxiliary graph on $V(G)$ containing exactly those edges of $G$ whose endpoints are not true twins. A vertex cover in this auxiliary graph is exactly a twin cover of $G$. Taking the endpoints of a maximal matching gives a twin cover $X$ of size $h\le2t$ in polynomial time.

Every outside clique has a neighbor in $X$, since $G$ is connected. Let $F_X$ be the missing edges inside $X$ and put $q=|F_X|\le\binom h2$. The graph $G+F_X$ is cop-win. This is because, $X$ is now a clique, and every vertex outside $X$ is dominated by any of its neighbors in $X$. We can delete all outside vertices and then dismantle $X$, thus getting a dismantling order.

If $k\ge q$, this construction gives a completion within the budget. Otherwise $k<q=O(t^2)$, so we apply Theorem~\ref{fpt_tc_budget_thm}, using $h\le2t$. The running time becomes
\[
 2^{O(k(h+\log(k+1)))}n^{O(1)}=2^{O(t^3)}n^{O(1)}.
\]
Testing the possible budgets from zero to $q$ also gives the optimum value and a corresponding completion in the same asymptotic time. Theorem~\ref{union_theorem} then combines the optima of the original components.
\end{proof}

\begin{corollary}\label{fpt_vc_cor}
\problem{Cop-Win Edge Completion} is FPT parameterized by $t=\vc(G)$ alone, in time $2^{O(t^3)}n^{O(1)}$.
\end{corollary}
\begin{proof}
Every vertex cover is a twin cover, so $\operatorname{tc}(G)\le\vc(G)$. Apply Theorem~\ref{fpt_tc_thm}.
\end{proof}

\subsection{Treewidth and the completion budget}\label{sec_tw}
The preceding algorithms used vertices or cliques that can be exchanged by automorphisms. For treewidth, we use a different method. A dismantling order can be described through an orientation of the graph, which allows us to apply a general algorithm for graphs of bounded treewidth.

For an orientation $D$ of $H$, write $N_D^+(v)$ for the outneighbors of $v$.
\begin{lemma}\label{orientation_lemma}
A nonempty graph $H$ is cop-win if and only if it has an acyclic orientation $D$ with a unique sink $r$ such that every $v\ne r$ has an outneighbor $w$ satisfying $N_D^+(v)\subseteq N_H[w].$
\end{lemma}
\begin{proof}
Take a dismantling order and orient each edge from its earlier (in the ordering) endpoint to its later endpoint. Clearly, the orientation is acyclic, and the last vertex is the unique sink. At the time a vertex $v$ is deleted, it has a later vertex $w$ that dominates it. Thus $w$ is an outneighbor of $v$ and sees every other outneighbor of $v$, giving the required inclusion.

Conversely, take a topological order of an orientation satisfying the condition. When $v\ne r$ is deleted, its remaining neighbors are exactly its outneighbors. The specified vertex $w$ is adjacent to $v$ and all of these remaining neighbors. Hence $w$ dominates $v$ at that step. The order dismantles $H$ to $r$.
\end{proof}

\begin{theorem}\label{fpt_tw_thm}
There is a computable function $f$ such that
\problem{Cop-Win Edge Completion} can be solved in
$f(k,\tw(G))(n+m)^{O(1)}$ time. A completion of size at most $k$
can be constructed whenever one exists.
\end{theorem}

\begin{proof}
We apply Courcelle's theorem for MSO$_2$ logic on graphs of
bounded treewidth~\cite{courcelle1990monadic}.
Let $t=\tw(G)$. Subdivide every edge
$uv$ once, replacing it with the path $u,e_{uv},v$, and denote
the resulting graph by $S(G)$. Mark original and subdivision
vertices by fixed unary predicates $\operatorname{Orig}$ and
$\operatorname{Sub}$. A width-$t$ tree decomposition of $G$
extends to $S(G)$ by attaching a bag $\{u,v,e_{uv}\}$ to a bag
containing $u,v$ for every original edge. Hence
\[
 \tw(S(G))\le\max\{t,2\},\qquad
 |V(S(G))|=n+m,\qquad |E(S(G))|=2m.
\]

Let $Z$ be an MSO$_2$ edge-set variable on $S(G)$. Require
exactly one of the two edges incident with each subdivision
vertex to belong to $Z$, and interpret its original endpoint
as the head of the corresponding original edge. Writing
$\operatorname{Sel}_Z(e,y)$ for the statement that $ey$ belongs
to $Z$, define
\[
 \begin{aligned}
 \operatorname{Arc}_Z(x,y)\equiv{}&
 \operatorname{Orig}(x)\land\operatorname{Orig}(y)\land x\ne y\\
 &{}\land\exists e\bigl[
 \operatorname{Sub}(e)\land
 \operatorname{Adj}_{S(G)}(e,x)\land
 \operatorname{Sel}_Z(e,y)\bigr].
 \end{aligned}
\]
Thus the permitted choices of $Z$ correspond exactly to
orientations of the original edges.

For each $j\in\{0,\ldots,k\}$, quantify $j$ ordered pairs
$(a_i,b_i)$ of original vertices. Require distinct, nonadjacent
endpoints in each pair and distinct unordered pairs. Different
pairs may share endpoints. The pair $(a_i,b_i)$ represents an
added edge oriented from $a_i$ to $b_i$. Adjacency and orientation
in the completed graph are described by
\[
 \begin{aligned}
 \operatorname{Arc}_j(x,y)&\equiv
 \operatorname{Arc}_Z(x,y)
 \lor\bigvee_{i=1}^j(x=a_i\land y=b_i),\\
 \operatorname{Adj}_j(x,y)&\equiv
 \operatorname{Adj}_G(x,y)
 \lor\bigvee_{i=1}^j
 \bigl((x=a_i\land y=b_i)\lor(x=b_i\land y=a_i)\bigr).
 \end{aligned}
\]
Here $\operatorname{Adj}_G(x,y)$ says that $x,y$ are distinct
original vertices with a common subdivision neighbor.
For $j=0$, the empty disjunctions are false.
These predicates are formula abbreviations. No candidate
nonedges are inserted into $S(G)$.

In the following conditions, all vertex variables and vertex
sets are restricted to original vertices using
$\operatorname{Orig}$. Require
\[
 \forall Y\subseteq V(G)\;
 \bigl(
 Y\ne\varnothing
 \Longrightarrow
 \exists v\in Y\;\forall u\in Y\;
 \neg\operatorname{Arc}_j(v,u)
 \bigr).
\]
This expresses acyclicity. Indeed, a directed cycle violates
the condition on its vertex set. Conversely, if every vertex
of a nonempty set has an outneighbor in that set, repeatedly
following outgoing edges produces a directed cycle.

Finally, require an original vertex $r$ with no outgoing edge
and, for every $v\ne r$, require
\[
 \exists w\,
 \bigl[
 \operatorname{Arc}_j(v,w)\land
 \forall u\,
 \bigl(
 \operatorname{Arc}_j(v,u)
 \Longrightarrow
 (u=w\lor\operatorname{Adj}_j(u,w))
 \bigr)
 \bigr].
\]
This says that $w$ is an outneighbor of $v$ satisfying
$N_D^+(v)\subseteq N_{G+F}[w]$. Every vertex other than $r$
has an outneighbor, so $r$ is the unique sink.

Let $\varphi_j$ assert the existence of $Z$, the endpoint pairs,
and $r$ satisfying these conditions. By
Lemma~\ref{orientation_lemma}, a satisfying assignment gives a
cop-win completion with exactly $j$ added edges. Conversely,
a cop-win completion with $j$ edges has an orientation from that
lemma, which supplies a satisfying assignment. Therefore
\[
 S(G)\models\bigvee_{j=0}^k\varphi_j
 \quad\Longleftrightarrow\quad
 \operatorname{copfill}(G)\le k.
\]

After expanding the abbreviations, the disjunction is an
MSO$_2$ sentence whose length depends only on $k$. Courcelle's
theorem~\cite{courcelle1990monadic}, together with standard FPT
tree-decomposition preprocessing, therefore gives running time
$f(k,t)(n+m)^{O(1)}$.

A completion can also be recovered by decision tests. First find
a successful $j$, and then fix the endpoint variables one at a
time, trying original vertices and retaining a choice that
admits an extension of all previously fixed choices.
Each choice is enforced by a singleton unary label, which
does not change treewidth. The formula length still depends
only on $k$. At most $(k+1)+2jn$ model-checking calls are sufficient,
so recovery remains within an FPT running time.
Note that the resulting pairs give the required fill set.
\end{proof}

The same result gives FPT for $k+\operatorname{fvs}(G)$, since deleting a feedback vertex set of size $s$ leaves a forest and $\tw(G)\le s+1$. The algorithm above does not settle the parameterization by treewidth alone, because it still depends on $k$.

\section{Conclusion and further questions}\label{section_conclusion}
Cop-win edge completion differs from chordal completion. It is $\mathrm{W}[2]$-hard for the fill budget, even on connected graphs with domination number two. The two lower bounds give exact algorithms for partial $2$-trees and block-cactus graphs, while the disjoint-union and join formulas give an algorithm for cographs. Complete bipartite graphs show that cop-win completion can need arbitrarily fewer edges than chordal completion.

The parameterized algorithms use symmetry for neighborhood diversity and twin cover, and an orientation argument for $k+\tw(G)$. The complexity for treewidth alone remains open, in particular at treewidth three. It would also be interesting to improve the dependence on twin cover and to find other classes where $\copfill(G)=\fillin(G)+c(G)-1$.

\paragraph*{Use of generative AI}
ChatGPT was used for literature searches, manuscript organization, and LaTeX preparation. It assisted with the triangle-space lower bound (Theorem~\ref{triangle_space_bound}) and the disjoint-union lower-bound argument (Lemma~\ref{cut_lemma} and Theorem~\ref{union_theorem}), and generated the treewidth-parameterized FPT algorithm (Section~\ref{sec_tw}). The author takes responsibility for the final manuscript and its mathematical correctness. 
\bibliographystyle{plainurl}
\bibliography{references}

\end{document}